# Multi-heterodyne differential absorption lidar based on dual-comb spectroscopy for simultaneous greenhouse gas and wind speed measurements

William Patiño Rosas[(a)] and Nicolas Cézard[(b)]

[(a)] ONERA/DOTA, Université Paris Saclay, F-91123 Palaiseau, France
[(b)] ONERA/DOTA, Université de Toulouse, F-31055 Toulouse France
william.patino@onera.fr

**Abstract:** In a recent study [1], we demonstrated a multi-heterodyne differential absorption lidar (DIAL) for greenhouse gas monitoring utilizing solid targets. The multi-frequency absorption measurement was achieved using electro-optic dual-comb spectroscopy (DCS) with a minimal number of comb teeth to maximize the optical power per tooth. In this work, we examine the lidar's working principle and architecture. Additionally, we present measurements of atmospheric $CO_2$ at 1572 nm over a 1.4 km optical path conducted in summer 2023. Furthermore, we show that, due to the high power per comb tooth, multi-frequency Doppler wind speed measurements can be performed using the DCS signal from aerosol backscattering. This enables simultaneous radial wind speed and path-average gas concentration measurements, offering promising prospects for novel concepts of tunable multi-frequency lidar systems capable of simultaneously monitoring wind speed and gas concentrations.

## 1. Introduction

Differential absorption lidar (DIAL) and open-path dual-comb spectroscopy (DCS) are two distinct approaches for the remote measurement of atmospheric gases using lasers. Both techniques have been successfully deployed for monitoring greenhouse gases, which are of special interest for environmental and climate change studies.

In typical DIAL systems, a small number (usually 2) of known and stabilized high-power laser wavelengths are emitted sequentially into the atmosphere over time. As they propagate, these wavelengths are differentially absorbed by the target gas according to their wavelength and are backscattered by aerosols in the atmosphere. When a solid target placed at the end of the line of sight reflects the laser signals, the technique is often referred to as integrated-path differential absorption (IPDA) lidar. In this case, we measure the path-average gas concentration. When the aerosol echo is utilized, range-resolved measurements become possible [2]. Additionally, detecting the aerosol backscattered signal in a heterodyne configuration allows for measuring the Doppler shift added to the particles and estimating wind speed along with gas concentration [3].

In open-path DCS, a pair of optical frequency combs (OFC) is transmitted through the atmosphere. Due to the comb nature of the signals, evenly-spaced wavelengths are emitted simultaneously, enabling high spectral resolution and facilitating multispecies gas sensing [4]. The number of wavelengths in open-path DCS can be very large (above $10^4$), resulting in smaller average power per wavelength compared to DIAL. To mitigate optical losses, a mirror is placed at the end of the line of sight to reflect the signal. The comb teeth are differentially absorbed by gases in the atmosphere. Finally, the multi-heterodyne interference signal of both OFCs, containing the optical absorption signature of the atmospheric column, is detected and processed. Configurations where only one comb propagates in the atmosphere have also been demonstrated [5].

In a previous study, we presented the development of a multi-heterodyne DIAL based on electro-optic DCS for greenhouse gas sensing [1]. The system utilizes a reduced number of comb teeth (3 or 5), combining the high power per wavelength characteristic of DIAL systems with the simultaneity and spectral regularity of DCS. This approach relaxes the frequency calibration constraints of DIAL systems and is robust to reflectance variations on natural targets, which can pose challenges for air and space-borne applications [6]. In this paper, we provide an overview of the system's operation and show an example of path-average $CO_2$ measurement conducted in summer 2023. Additionally, we demonstrate that, due to the high power per wavelength, the aerosol backscattering signal can be leveraged to estimate radial wind speed in addition to the path-average gas concentration. These results open the door for new fiber-based, multi-frequency lidar architectures capable of simultaneously monitoring gas concentration and wind speed.

## 2. Multi-heterodyne DIAL based on DCS

Fig. 1 presents a schematic representation of the lidar setup. A fiber-coupled laser diode (LD) generates two OFCs through pure electro-optic phase modulation. An acousto-optic modulator (AOM) operates in a pulsed regime, introducing a frequency shift of 40 MHz to one of the OFCs and generating nearly Gaussian pulses of 1 µs, which are subsequently amplified by an Erbium-doped fiber amplifier (EDFA). The pulses are collimated and expanded thanks to a telescope, and emission and reception are separated by a polarization beam splitter (PBS), facilitated by the insertion of a quarter-wave plate (QWP) between the telescope's lenses. Further details regarding the lidar architecture and the inversion procedure can be found in [1].

For $CO_2$ sensing, three wavelengths centered at 1572.02 nm were generated, with a frequency spacing of 2500 MHz. Following amplification, we obtained 20 µJ pulses at a pulse repetition frequency (PRF) of 20 kHz. The second OFC had a frequency spacing of 2507.5 MHz and was dedicated to DCS. Consequently, the resulting radio frequency (RF) comb was centered at 40 MHz with a tooth spacing of 7.5 MHz.

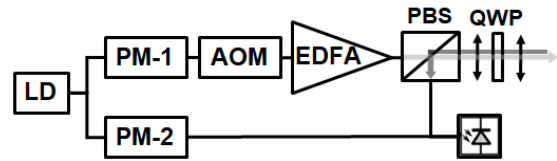

**Figure 1.** Lidar setup. LD, laser diode; PM, phase modulator; AOM, acousto-optic modulator; EDFA, Erbium-doped fiber amplifier; PBS, polarization beam splitter; QWP, quarter-wave plate.

Another detector, not depicted in Fig. 1, receives a reference DCS signal, which results from the interference of the two OFCs before interaction with the atmosphere. The DCS signals are divided into overlapping 1 µs segments, which are then Fourier transformed. The power spectra from different range gates are averaged, and the noise level is subsequently subtracted to generate a power spectrogram as a function of distance. For the IPDA measurement, only the range gate corresponding to the solid target return is utilized. The ratio between the solid target return power spectrum and the reference power spectrum gives the total differential transmission of the atmospheric column at the comb wavelengths, which is further utilized to estimate the path-average gas concentration using a maximum likelihood estimator based on the known line shape provided by the High-Resolution Transmission Molecular Absorption Database HITRAN-2023.

## 3. Simultaneous atmospheric $CO_2$ and wind speed measurements

The system was deployed at the ONERA site in the south of Toulouse, France, and IPDA measurements of atmospheric $CO_2$ were conducted using a solid target located 700 m away, as explained in [1]. Fig. 2 displays the measurement power spectrogram as a function of distance. A parasitic reflection, characteristic of the monostatic lidar architecture, is observed, along with the solid target return at 700 m. The color scale of the spectrogram has been adjusted to highlight the presence of aerosol backscattering.

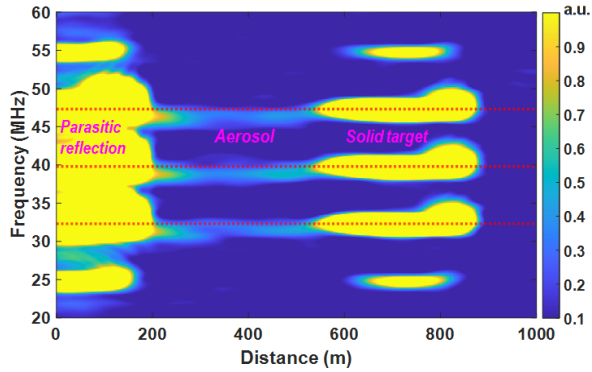

**Figure 2.** Power spectrogram as a function of the distance for the IPDA measurement of atmospheric $CO_2$. The contrast has been adjusted to appreciate aerosol backscattering.

Following the outlined measurement procedure, we estimated the average $CO_2$ concentration over several hours with a precision of approximately 5%, mainly limited by speckle noise. Fig. 3(a) shows the estimated $CO_2$ concentration as a function of local time. To smooth the measurements, a moving average over $1.2 \times 10^6$ pulses was applied. The inset figure depicts the positioning of the OFC relative to the absorption line.

The presence of aerosol backscattering enables the estimation of the Doppler shift induced by wind speed. A closer examination of Fig. 2 reveals that, for this particular spectrogram, the aerosol signal is shifted by approximately -1.5 MHz relative to the solid target return, indicated by the dashed lines. Given that the frequency spacing of the different comb teeth is very small compared to their central frequency, the Doppler shift is approximately the same for all three wavelengths. To estimate the radial wind speed, we calculate the frequency barycenter for each of the wavelengths as a function of distance, following a standard procedure in coherent Doppler wind lidar (CDWL) [7]. Subsequently, we determine the weighted mean Doppler shift, with the weights determined by the powers of the comb teeth, which ultimately yields the radial wind speed. Moreover, the standard deviation of the individual Doppler shifts provides insight into the measurement precision.

Previous works have already explored electro-optic modulation to generate evenly spaced frequencies that can be used for multi-wavelength DIAL [2] or to increase the detection range of CDWLs [8]. To the best of our knowledge, this is the first demonstration of a lidar system using DCS to simultaneously measure gas concentrations and wind speed.

Fig. 3(b) illustrates the estimated radial wind over time as a function of distance. Each point corresponds to an average of $1.5 \times 10^5$ pulses. The spatial resolution (150 m) of the measurement is determined by the pulse duration and is analyzed using a moving window with a 90% overlap. Note that the wind speeds at the solid target and parasitic reflection are centered around 0 m/s, indicating no important bias in the measurements, except for a slight frequency chirp observed at the end of the pulse, possibly originating during pulsed amplification [9]. The availability of measurements is contingent upon aerosol concentrations. Furthermore, the power of individual heterodyne beatings is reduced by a factor of approximately 9 (arising from 3 comb teeth for each OFC) compared to a single-frequency CDWL under the same measurement conditions [1]. Consequently, for some measurements, the signal power was insufficient to precisely determine the wind speed. We applied a mask to display only those measurements where the standard deviation of the estimated wind speed is below 0.3 m/s. However, it is noteworthy that for the majority of measurements, the estimated radial wind standard deviation is better than 0.3 m/s, indicating the system's capability to monitor wind speed along with path-average gas concentrations.

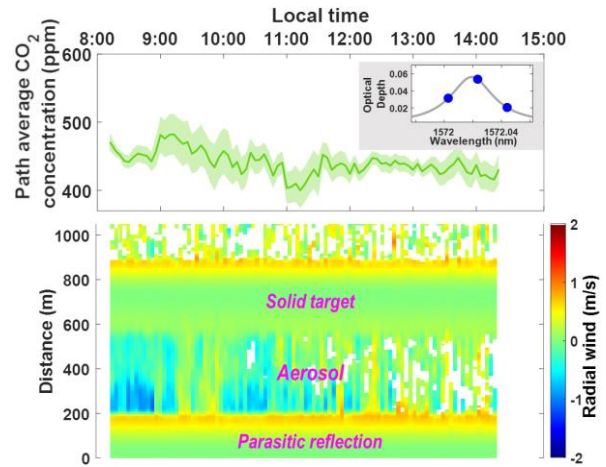

**Figure 3.** (a) Average $CO_2$ concentration over time obtained from the IPDA measurement. The inset figure depicts the positioning of the comb relative to the absorption line. (b) Estimated radial wind speed as a function of distance. A mask is applied to display only the measurements with a standard deviation below 0.3 m/s.

## 4. Conclusion and perspectives

We have presented a multi-heterodyne DIAL lidar for monitoring $CO_2$ atmospheric concentrations, based on DCS with OFC produced by electro-optic phase modulation. Utilizing three comb wavelengths centered around 1572.02 nm, in pulsed mode, with 1 µs/20 µJ pulses and a collimated beam, we measured the path-average $CO_2$ concentrations with a precision of 5% using the lidar echo from a solid target placed 700 m from the emitter. Additionally, we highlighted the presence of aerosol backscattering, which was subsequently used to estimate the radial wind speed using a simple processing method. For most of the measurements, the standard deviation in the measured wind speed was below 0.3 m/s, demonstrating the capability of the system to simultaneously monitor wind speed and gas concentrations. Although we were able to observe aerosol backscattering, its power was insufficient to conduct range-resolved DIAL. Indeed, the chosen $CO_2$ absorption line has a relatively low absorption coefficient (less than 0.1/km), such that conducting range-resolved DIAL measurements over optical paths below 1 km requires very high precision in the differential absorption measurement.

Perspectives for this work include, on one hand, increasing the laser power. Specifically, when working with fiber amplifiers, the use of OFCs allows for mitigating stimulated Brillouin scattering and achieving higher peak powers [10, 11]. This would enable an increase in the number of wavelengths, providing higher spectral sampling of the absorption line and allowing for the measurement of path-average concentrations over longer atmospheric columns. Additionally, focusing the laser would increase the magnitude of the aerosol signal, potentially enabling simultaneous range-resolved DIAL and wind speed measurements. On the other hand, the signal processing method used to estimate the wind speed is simple and does not fully exploit the comb nature of the DCS signals. More sophisticated signal processing algorithms could be developed to enhance wind speed measurement availability and precision.